\documentclass[jkps,showpacs,showkeys]{revtex4}
\usepackage[dvips]{graphicx}
\usepackage{latexsym}

\begin{document}
%\preprint{\vbox{\hbox{\tt gr-qc/0703065}}}

\title[]{Positive Equation-of-state Parameter in the Accelerating
  Dilaton Cosmology}
\author{Wontae \surname{Kim}}
  \email{wtkim@sogang.ac.kr}
  \affiliation{Department of Physics and Center for Quantum Spacetime, \\
  Sogang University, C.P.O. Box 1142, Seoul 100-611, Korea}
\author{Myung Seok \surname{Yoon}}
  \email{younms@sogong.ac.kr}
  \affiliation{Center for Quantum Spacetime, Sogang University, Seoul
    121-742, Korea}

\date{\today}

%\bigskip
\begin{abstract}
In a semiclassically quantized two-dimensional cosmological model, it
can be shown that the parameter of the equation of state for the
accelerating universe can be positive due to the negative energy
density and the negative pressure, which is a little different from
the conventional wisdom that the parameter is negative with the
positivity of the energy density. Furthermore, we show that the full
parameter composed of the classical and the quantum-mechanical
contributions is positive and finite even though the partial state
parameter from the quantum-mechanical contribution is not positive
definite, which  means that the state parameter is not perturbatively
additive in this model.
\end{abstract}


\maketitle
%\bigskip

\section{Introduction}
\label{sec:intro} 

Recently, much attention has been paid to the accelerating expansion
of the universe\cite{perlmutter,turner,spergel}, which is essentially
related to the equation-of-state
parameter\cite{caldwell,kkkkl,kim}. In the Einstein gravity, the
decelerating universe satisfying the positive energy conditions
appears, and the energy density and the pressure are naturally
positive definite. However, the dark energy related to the
accelerating expansion of the universe is defined by the negative
state parameter\cite{carroll,at,kobayashi,fjt,df,myung:cai}, which
means that the pressure is negative.
  
In the two-dimensional gravity\cite{cghs,str,rst,mrcm,bpp,kv,bc,bk:1,ksy},
it is a little easier to treat some unsolved cosmological
problems\cite{gv,rey,bk,ky:dg,ky:bd,ky:2d,cdkz}, so the cosmological
solutions describing the phase change from the decelerating universe
to the accelerating universe are obtained\cite{ky:ncdc} by assuming 
noncommutativity among the relevant fields\cite{sw,bn,vas,ko}.  In
this toy model, the equation of state is based on the
matter part having been solved in a self-consistent manner.  However,
in the semi-classical quantized region, the fields are redefined,
and the newly defined fields are a combination of the metric and
dilaton, so that, contrary to the conventional Friedman
equation, it is not straightforward to see the relation between the
acceleration for the scale factor and the energy density along with
the pressure. Therefore, it would be interesting to study whether the
energy density and the pressure affect the equation-of-state
parameter or not because the quantum-mechanically-induced energy
density and pressure may modify the signature for the state
parameter.
  
In this paper, we would like to calculate the classical and the
quantum-mechanically-induced energy-momentum tensors separately in
comoving coordinates in order to find the signature of the
equation-of-state parameter in this phase changing from the
decelerating universe to the accelerating one. In Sec.~\ref{sec:gen},
the stress-energy-momentum tensors are formally calculated in this 
dilaton cosmology, and their expressions for the energy density and the
pressure are written in the form of a perfect fluid. In
Sec.~\ref{sec:energy}, using the noncommutative algebra during finite
time, we obtain the cosmological solution describing the accelerating 
universe from the decelerating phase; then, this geometry
is patched up by regular geometry to avoid the future curvature
singularity. In fact, if noncommutativity between fields is
applied to all cosmic time, then the future curvature
singularity appears in a finite proper time. Next, the energy
density and the pressure as a perfect fluid are investigated in this
geometry. Then, we obtain the equation-of-state parameters for
the classical and the quantum-mechanical cases separately to clarify
their roles and their signatures. Consequently, the
total energy density and the pressure derived from the classical and
the quantum mechaninical energy densities and the pressures give the
so-called total state parameter, which is curiously always positive
definite, because the total energy density is negative, as is total
pressure. Finally, a summary and discussion are 
given in Sec.~\ref{sec:dis}.
 
%%%%%%%%%%%%%%%%%%%%%  %%%%%%%%%%%%%%%%%%%
\section{Energy-momentum tensors in the dilaton cosmology}
\label{sec:gen}

We now start with the following Callan-Giddings-Harvey-Strominger
(CGHS) model\cite{cghs}, which is split into two pieces in order to
regard the kinetic part of the dilaton as matter. Thus, the action can
be written as
\begin{equation}
  \label{S}
   S = S_{\rm G} + S_{\rm Cl} + S_{\rm Qt}.
\end{equation}
The gravitational action is defined by
\begin{equation}
  \label{S:G}
  S_{\rm G} = \frac{1}{2\pi} \int d^2 x \sqrt{-g} e^{-2\phi} R,
\end{equation}
which is trivial in its form without additional terms because the
spacetime is flat. The remaining  classical matter and its quantum
correction are given in the form
\begin{eqnarray}
  S_{\rm Cl} &=& \frac{1}{2\pi} \int d^2 x \sqrt{-g} \left[ e^{-2\phi}
    \left( 4 (\nabla \phi)^2 + 4 \lambda^2\right) - \frac12
     \sum_{i=1}^{N} (\nabla f_i)^2 \right], \label{S:cl} \\
  S_{\rm Qt} &=& \frac{\kappa}{2\pi} \int \sqrt{-g} \left[ 
     - \frac14 R\frac{1}{\Box} R + (\nabla\phi)^2 - \phi R \right],
     \label{S:qt}
\end{eqnarray}
where $\kappa=(N-24)/12$ and the cosmological constant $\lambda^2$
will be set to zero. The nonlocal form of the action in
Eq.~(\ref{S:qt}) is written, by introducing an auxiliary field $\psi$
for later convenience, as
\begin{equation}
  \label{S:qt:nonlocal}
  S_{\rm Qt} = \frac{\kappa}{2\pi}\int d^2x\/ \sqrt{-g}\left[
    \frac14 R\psi-\frac{1}{16}(\nabla\psi)^2 +
    \left(\nabla\phi\right)^2 - \phi R\right].
\end{equation}
The total stress-energy-momentum tensor is given as
\begin{equation}
  \label{T}
  T_{\mu\nu}^{\rm M} = T_{\mu\nu}^{\rm Cl} + T_{\mu\nu}^{\rm Qt},
\end{equation}
where the energy-momentum tensors of the classical and the quantum
matter are defined as
\begin{eqnarray}
  T_{\mu\nu}^{\rm Cl} &\equiv& - \frac{2\pi}{\sqrt{-g}} 
         \frac{\delta S_{\rm Cl}}{\delta g^{\mu\nu}} \nonumber \\
     &=& e^{-2\phi} \left[ 2g_{\mu\nu} (\nabla\phi)^2 - 
         4\nabla_\mu \phi \nabla_\nu \phi \right] + \frac12
       \sum_{i=1}^N \left[\nabla_\mu f_i \nabla_\nu 
         f_i -\frac{1}{2}(\nabla f_i)^2 \right], 
         \label{T:cl} \\
  T_{\mu\nu}^{\rm Qt} &\equiv& - \frac{2\pi}{\sqrt{-g}} 
         \frac{\delta S_{\rm Qt}}{\delta g^{\mu\nu}} \nonumber \\
     &=& \frac{\kappa}{4}\left[\nabla_\mu\nabla_\nu\psi +
         \frac14\nabla_\mu\psi\nabla_\nu\psi -
         g_{\mu\nu}\left(\Box\psi+\frac18(\nabla\psi)^2 \right)\right]
         \nonumber \\
     & & -\kappa\bigg[
         \nabla_\mu\nabla_\nu\phi - g_{\mu\nu}\Box\phi\bigg]
         - \kappa\left[\nabla_\mu\phi\nabla_\nu\phi
         - \frac12 g_{\mu\nu}(\nabla\phi)^2\right],
         \label{T:qt}
\end{eqnarray}
respectively. We would like to investigate the effect of quantum
matter on the expansion of universe. Thus, for simplicity, we assume
that there are no conformal matter fields, \textit{i.e.}, $f_i =
0$. Then, the equation of motion for $\psi$ is easily obtained as
$\Box\psi = -2R$. In the conformal gauge, $ds^2 = -e^{2{\tilde\rho}}
dx^+ dx^- = -e^{2{\tilde\rho}(t)} (dt^2 - dx^2)$, the
stress-energy-momentum tensors in Eqs.~(\ref{T:cl}) and (\ref{T:qt})
are written as
\begin{eqnarray}
  T_{\pm\pm}^{\rm Cl} &=& -4 e^{-2\phi} (\partial_\pm \phi)^2 \nonumber \\
     &=& - e^{-2\phi} \dot\phi^2, \label{T++:cl} \\
  T_{+-}^{\rm Cl} &=& -2 e^{-2\phi} \partial_+ \phi
         \partial_- \phi \nonumber \\
     &=& - \frac12 e^{-2\phi} \dot\phi^2, \label{T+-:cl} \\
  T_{\pm\pm}^{\rm Qt} &=& \kappa\left[\partial_\pm^2{\tilde\rho}
        - (\partial_\pm{\tilde\rho})^2 - t_\pm(x^\pm)\right]
        - \kappa\left[\partial_\pm^2\phi
        - 2\partial_\pm{\tilde\rho}\partial_\pm\phi\right]
        - \kappa(\partial_\pm\phi)^2 \nonumber \\
     &=& -\kappa t_{\pm} + \frac{\kappa}{4}(\ddot{\tilde\rho} -
     \ddot\phi) - \frac{\kappa}{4} (\dot{\tilde\rho}-\dot\phi)^2,
       \label{T++:qt} \\
  T_{+-}^{\rm Qt} &=& -\kappa\partial_+\partial_-{\tilde\rho} +\kappa
         \partial_+\partial_- \phi \nonumber \\
     &=& -\frac{\kappa}{4} (\ddot{\tilde\rho} - \ddot\phi),
         \label{T+-:qt}
\end{eqnarray}
where $t_{\pm}$ reflects the nonlocality of the induced gravity of the
conformal anomaly and the overdot denotes the derivative with respect
to $t$. We obtained these expressions by substituting the solution for
the auxiliary field $\psi$ from the equation of motion for $\psi$.

In a comoving coordinate system, $ds^2 = -d\tau^2 + a^2(\tau) dx^2$,
the stress-energy-momentum tensors of a perfect fluid are given as
\begin{equation}
  \label{T:perfect}
  \hat{T}_{\mu\nu}^{\rm M} = p g_{\mu\nu} + (p + \rho) u_\mu u_\nu,
\end{equation}
where  $u_\mu=(1,0)$ and $p$ and $\rho$ are the pressure and the
energy density, respectively. Then, we obtain the expressions for the
energy density and the pressure as
\begin{eqnarray}
  \rho &=& \hat{T}_{\tau\tau}^M, \label{rho:def} \\
  p &=& \frac{1}{a^2} \hat{T}_{xx}^M, \label{p:def}
\end{eqnarray}
which are composed of the classical and the quantum matter as follows:
\begin{eqnarray}
  \rho &=& \rho_{\rm Cl} + \rho_{\rm Qt}, \label{rho:non} \\
  p &=& p_{\rm Cl} + p_{\rm Qt}. \label{p:non}
\end{eqnarray}
By a coordinate transformation, the relations between the
stress-energy-momentum tensors are obtained as 
\begin{eqnarray}
  \hat{T}_{\tau\tau}^{\rm M}(\tau) &=& e^{-2{\tilde\rho}} (T_{++}^{\rm M} +
      2 T_{+-}^{\rm M} + T_{--}^{\rm M}), \label{T:tautau} \\
  \hat{T}_{\tau x}^{\rm M} (\tau) &=& e^{-{\tilde\rho}} (T_{++}^{\rm M} -
      T_{--}^{\rm M} ), \label{T:taux}\\
  \hat{T}_{xx}^{\rm M} &=& T_{++}^{\rm M} - 2 T_{+-}^{\rm M} +
      T_{--}^{\rm M}, \label{T:xx}
\end{eqnarray}
with the comoving time $\tau$, where $\tau = \int e^{{\tilde\rho}(t)} dt$.

%%%%%%%%%%%% t_1 < t < t_2   %%%%%%%%%%%

\section{The distribution of the energy density and the pressure}
\label{sec:energy}

In the conformal gauge, $ds^2 = -e^{2\rho} dx^+ dx^-$, without
conformal fields, $f_i=0$, the total action and the constraint
equations are written as
\begin{equation}
 S = \frac{1}{\pi} \int\/ d^2 x \left[ e^{-2\phi} \left(
       2\partial_+\partial_-\rho - 4\partial_+\phi\partial_-\phi
       \right) - \kappa\big( \partial_+\rho\partial_-\rho
       + 2\phi\partial_+\partial_-\rho 
     + \partial_+\phi\partial_-\phi \big) \right] \label{act:conf}
\end{equation}
and
\begin{eqnarray}
  & & e^{-2\phi}\left[ 4\partial_\pm\rho\partial_\pm\phi
      - 2\partial_\pm^2\phi \right]
      + \kappa\left[ \partial_\pm^2\rho
      - \left(\partial_\pm\rho\right)^2\right] \nonumber \\
  & & \qquad\qquad\qquad - \kappa\left(
      \partial_\pm^2\phi - 2\partial_\pm\rho\partial_\pm\phi\right)
      - \kappa\left(\partial_\pm\phi\right)^2
      - \kappa t_\pm = 0. \label{constr:conf}
\end{eqnarray}
Defining new fields as $\Omega = e^{-2\phi}$ and $\chi = \kappa (\rho
-\phi) + e^{-2\phi}$\cite{bpp,bc}, the gauge fixed action is obtained
in the simplest form of
\begin{equation}\label{act:new}
S = \frac{1}{\pi} \int\/d^2 x \left[
    \frac{1}{\kappa}\partial_+\Omega\partial_-\Omega 
    - \frac{1}{\kappa}\partial_+\chi\partial_-\chi \right]
\end{equation}
and the constraints are given by
\begin{equation}
\kappa t_\pm = \frac{1}{\kappa}\left( \partial_\pm\Omega\right)^2 
    - \frac{1}{\kappa} \left(\partial_\pm\chi\right)^2 
    + \partial_\pm^2\chi. \label{constr:new}
\end{equation}

In a homogeneous spacetime, the Lagrangian and the constraints are
obtained as
\begin{eqnarray}
  L &=& \frac{1}{4\kappa} \dot{\Omega}^2 
        - \frac{1}{4\kappa} \dot{\chi}^2, \label{L} \\
  \frac{1}{4\kappa} \dot{\Omega}^2 \!&-&\! \frac{1}{4\kappa}
        \dot{\chi}^2 + \frac14 \ddot{\chi} - \kappa t_{\pm}=0, 
        \label{con}
\end{eqnarray}
where the action is redefined by $S/L_0 = \frac{1}{\pi} \int dt L$ and
$L_0=\int dx$, and the overdot denotes the derivative with respect to
the cosmic time $t$. Then, the Hamiltonian becomes
\begin{equation}
  \label{H}
  H = \kappa P_\Omega^2 - \kappa P_\chi^2
\end{equation}
in terms of the canonical momenta $P_\chi = - \frac{1}{2\kappa}
\dot{\chi}$ and $P_\Omega = \frac{1}{2\kappa} \dot{\Omega}$.

We consider the modified Poisson brackets corresponding to the
noncommutative algebra in the quantized theory\cite{sw,bn}:
\begin{eqnarray}
  & & \{ \Omega, P_\Omega \}_{\mathrm{MPB}} = 
      \{ \chi, P_\chi \}_{\mathrm{MPB}} = 1, \nonumber \\ 
  & & \{ \chi, \Omega \}_{\mathrm{MPB}} = 
      \theta_1 [\epsilon(t-t_1) - \epsilon(t-t_2)], \nonumber \\
  & & \{ P_\chi, P_\Omega \}_{\mathrm{MPB}} = 
      \theta_2 [\epsilon(t-t_1) - \epsilon(t-t_2)], \nonumber\\
  & & \mathrm{others} = 0, \label{PB:non}
\end{eqnarray}
where $\theta_1$ and $\theta_2$ are two independent positive
constants, and $\epsilon(t)$ is a step function, $1$ for $t>0$ and $0$
for $t<0$. Thus, these are nontrivial and $\theta$-dependent for the
finite time interval $t_1 < t < t_2$. With the
Hamiltonian in Eq.~(\ref{H}), the equations of motion are obtained as
\begin{eqnarray}
  & & \dot\chi = \{ \chi, H \}_{\mathrm{MPB}} = -2\kappa P_\chi, \quad
      \dot\Omega =  \{ \Omega, H \}_{\mathrm{MPB}} = 2\kappa P_\Omega, 
      \label{1st:x_non}\\
  & & \dot{P}_\chi =  \{ P_\chi, H \}_{\mathrm{MPB}} =2\kappa \theta_2
      P_\Omega, \quad \dot{P}_\Omega = \{ P_\Omega, H \}_{\mathrm{MPB}} 
      = 2\kappa\theta_2 P_\chi. \label{1st:p_non}
\end{eqnarray}
Note that the momenta are no longer constants of the motion because of
the nonvanishing $\theta_2$; hence, a new set of equations of motion
are obtained from Eqs.~(\ref{1st:x_non}) and (\ref{1st:p_non}): 
\begin{equation}
  \label{eom:non}
  \ddot\chi = -2\kappa \theta_2 \dot\Omega, \quad \ddot\Omega =
  -2\kappa \theta_2 \dot\chi.
\end{equation}
The solutions for the above coupled equations of motion are easily
obtained as
\begin{eqnarray}
  \Omega &=& \alpha e^{2\kappa\theta_2 t} + \beta
            e^{-2\kappa\theta_2 t} + A, \label{sol_gen:Omega_non} \\
  \chi &=& \alpha e^{2\kappa\theta_2 t} - \beta
            e^{-2\kappa\theta_2 t} + B, \label{sol_gen:chi_non}
\end{eqnarray}
where $\alpha$, $\beta$, $A$, and $B$ are constants, and they should
satisfy the constraint equation in Eq.~(\ref{con}),
\begin{equation}
  \label{constr_gen:non}
  \kappa t_\pm = \kappa^2 \theta_2^2 (\alpha e^{2\kappa\theta_2 t} 
      -\beta e^{-2\kappa\theta_2 t}) -4\kappa\theta_2^2 \alpha\beta,
\end{equation}
which determines the unknown time-dependent function $t_\pm$. 

Now, for $\beta=-\alpha>0$, the solutions and the constraints
are written as
\begin{eqnarray}
  \Omega &=& e^{-2\phi} = 2 \beta \sinh (2\kappa\theta_2 t) + A,
     \label{sol:Omega_non} \\
  \chi &=& \kappa(\rho - \phi) + e^{-2\phi} =  - 2 \beta \cosh
     (2\kappa\theta_2 t) + B, \label{sol:chi_non}
\end{eqnarray}
and
\begin{equation}
  \label{constr:non}
  \kappa t_{\pm} = - 2 \beta \kappa^2 \theta_2^2 \cosh (2\kappa\theta_2
  t) + 4\kappa\beta^2\theta_2^2.
\end{equation}
Note that among the two positive constants, only
$\theta_2$ plays an important role in our analysis. Furthermore,
$\Omega = e^{-2\phi}$ in Eq.~(\ref{sol:Omega_non}) is positive
definite, so the initial time should be restricted to $t_1 > -
[1/(2\kappa\theta_2)] \sinh^{-1} [A/(2\beta)]$. Especially, for $A=0$,
the time interval become $0<t_1 < t < t_2$. Hereafter, we regard $t_1$
as  the initial time of the beginning of the universe in our model. 
Since $da(\tau)/d\tau = d{\tilde\rho}(t)/dt$, we obtain the expanding
velocity of the universe as
\begin{equation}
  \label{vel:non}
  \frac{da(\tau)}{d\tau} = - 2\beta\theta_2 \left[ \cosh
   (2\kappa\theta_2 t) + \sinh (2\kappa\theta_2 t) + 
   \frac{\kappa \cosh (2\kappa\theta_2
   t)}{A+2\beta\sinh(2\kappa\theta_2 t)} \right],
\end{equation}
which is always negative since $\cosh (2\kappa\theta_2 t) > |\sinh
(2\kappa\theta_2 t) |$ and $\Omega = A+ 2\beta \sinh (2\kappa\theta_2
t) >0$ for any time $t$. 

At this juncture, from Eqs.~(\ref{sol:Omega_non}) and
(\ref{sol:chi_non}), we calculate the curvature scalar related to the
acceleration and the deceleration in terms of $R=2\ddot{a}/a$ in
comoving coordinates; then, 
\begin{eqnarray}
  R_{\theta} &=& -8 \beta \kappa^2\theta_2^2 \frac{\exp(4\beta %
        \cosh(2\kappa\theta_2 t) -2B)}{2\beta \sinh (2\kappa %
        \theta_2 t) + A} \bigg[ \cosh(2\kappa\theta_2 t) \left(
        2\beta e^{2\kappa\theta_2 t} + A \right) \nonumber \\
  & &  - \frac{2}{\kappa} e^{2\kappa\theta_2 t} \left(
        2\beta\sinh(2\kappa\theta_2 t) + A\right) \left(
        2\beta\sinh(2\kappa\theta_2 t) + A + \frac{\kappa}{2}\right)
        \bigg]. \label{R:non}
\end{eqnarray}
By substituting the solutions in Eqs.~(\ref{sol:Omega_non}) and
(\ref{sol:chi_non}) into Eqs.~(\ref{T++:cl}), (\ref{T+-:cl}),
(\ref{T++:qt}), and (\ref{T+-:qt}), the stress-energy-momentum tensors
are calculated in the conformal gauge as
\begin{eqnarray}
  T_{\pm\pm}^{\rm Cl} &=& - \frac{4\beta^2 \kappa^2 \theta_2^2%
     \cosh^2(2\kappa\theta_2 t)}{2\beta\sinh (2\kappa\theta_2 t)%
     + A}, \label{T++:cl:non} \\
  T_{+-}^{\rm Cl} &=& - \frac{2\beta^2 \kappa^2 \theta_2^2%
     \cosh^2(2\kappa\theta_2 t)}{2\beta\sinh (2\kappa\theta_2 t)%
     + A}, \label{T+-:cl:non} \\
  T_{\pm\pm}^{\rm Qt} &=& -\kappa t_\pm - 2\beta \kappa^2 \theta_2^2
    e^{2\kappa\theta_2 t} - 4\beta^2\kappa\theta_2^2
    e^{4\kappa\theta_2 t}, \label{T++:qt:non} \\
  T_{+-}^{\rm Qt} &=& 2\beta\kappa^2\theta_2^2 e^{2\kappa\theta_2 t}. 
    \label{T+-:qt:non}
\end{eqnarray}
Then, by using Eqs.~(\ref{rho:def}) and (\ref{p:def}), we obtain the
energy densities and the pressures of the classical and the quantum
matter as
\begin{eqnarray}
  \rho_{\rm Cl} &=& - 12 \beta^2 \kappa^2 \theta_2^2 \cosh^2
      (2\kappa\theta_2 t) \exp \left[ \frac{2}{\kappa}(A-B) +
      \frac{4\beta}{\kappa} e^{2\kappa\theta_2 t} \right],
      \label{rho:cl:non} \\
  \rho_{\rm Qt} &=& (2\beta \sinh (2\kappa\theta_2 t) + A) \exp \left[
      \frac{2}{\kappa}(A-B) + \frac{4\beta}{\kappa}
      e^{2\kappa\theta_2 t} \right] \nonumber \\
    & & \times\left( -8\kappa\beta^2 \theta_2^2
      e^{4\kappa\theta_2 t} - \kappa(t_+ + t_-) \right),
      \label{rho:qt:non} \\
  p_{\rm Cl} &=& - 4 \beta^2 \kappa^2 \theta_2^2 \cosh^2
      (2\kappa\theta_2 t) \exp \left[ \frac{2}{\kappa}(A-B) +
      \frac{4\beta}{\kappa} e^{2\kappa\theta_2 t} \right],
      \label{p:cl:non} \\
  p_{\rm Qt} &=& (2\beta \sinh (2\kappa\theta_2 t) + A) \exp \left[
      \frac{2}{\kappa}(A-B) + \frac{4\beta}{\kappa}
      e^{2\kappa\theta_2 t} \right] \nonumber \\
    & & \times \left( -8\kappa\beta \theta_2^2
      e^{2\kappa\theta_2 t} \left( \kappa + \beta e^{2\kappa\theta_2
      t}\right) - \kappa(t_+ + t_-) \right).
      \label{p:qt:non}
\end{eqnarray}

%\subsection{Commutative Cosmology: $t>t_2$}

For $t>t_2$, the conventional Poisson brackets are recovered as follows:
\begin{equation}
  \label{PB:com}
  \{\Omega, P_\Omega\}_{\mathrm{PB}} = \{\chi, P_\chi\}_{\mathrm{PB}} =1,
  \quad \mathrm{others} = 0,
\end{equation}
and the Hamiltonian equations of the motion in Ref.\cite{bk:1} are given
by $\dot {\cal O}  = \{ {\cal O}, H \}_{\mathrm{PB}}$, where ${\cal O}$
represents fields and corresponding momenta. They are explicitly
written as 
\begin{eqnarray}
   & & \dot\chi = - 2\kappa  P_\chi, \quad \dot\Omega = 2\kappa
       P_\Omega, \label{1st:x_com} \\
   & & \dot{P}_\chi =0, \quad \dot{P}_\Omega = 0. \label{1st:p_com}
\end{eqnarray}
Since the momenta $P_\Omega$ and $P_\chi$ are constants of the motion as
seen from Eq.~(\ref{1st:p_com}), we easily obtain the solutions as 
\begin{eqnarray}
  & & \Omega =  2\kappa P_{\Omega_0} t + A_0, 
      \label{sol:Omega_com} \\
  & & \chi =  -2\kappa P_{\chi_0} t + B_0,
      \label{sol:chi_com}
\end{eqnarray}
where $ P_\Omega = P_{\Omega_0}$, $P_\chi = P_{\chi_0}$, and $A_0$ and
$B_0$ are arbitrary constants. Next, the dynamical solutions in
Eqs.~(\ref{sol:Omega_com}) and (\ref{sol:chi_com}) should satisfy the
constraint in Eq.~(\ref{con}),
\begin{equation}
  \label{constr:com}
  \kappa t_\pm = \kappa (P_{\Omega_0}^2 - P_{\chi_0}^2),  %% constraints
\end{equation}

On the other hand, by using Eqs. (\ref{sol:Omega_com}) and
(\ref{sol:chi_com}), the curvature scalar is calculated as
\begin{equation}
  \label{R:com}
  R = 4\kappa^2 P_{\Omega_0}^2 e^{-2\rho + 4\phi} 
    = 4 \kappa^2 P_{\Omega_0}^2 \frac{e^{ -2B_0 +4 \kappa P_{\chi_0} 
      t} }{ A_0 + 2\kappa P_{\Omega_0} t }. 
\end{equation}
Since $\Omega=e^{-2\phi}$ in Eq.~(\ref{sol:Omega_com}) should be
positive definite, there are two types of branches: The first one is
$t>-A_0/(2\kappa P_{\Omega0})$ for the positive charge of
$P_{\Omega_0}>0$, and the second is $t<A_0/(2\kappa P_{\Omega0})$
for the negative charge of $P_{\Omega_0}<0$. Note that the universe is
always accelerating irrespective of the vacuum energy density.

By substituting the solutions in Eqs.~(\ref{sol:Omega_com}) and
(\ref{sol:chi_com}) into Eqs.~(\ref{T++:cl}), (\ref{T+-:cl}),
(\ref{T++:qt}), and (\ref{T+-:qt}), we obtain the
stress-energy-momentum tensor as
\begin{eqnarray}
  T_{\pm\pm}^{\rm Cl} &=& - \frac{\kappa^2 P_{\Omega_0}^2}{2\kappa%
    P_{\Omega_0} t + A_0}, \label{T++:cl:com} \\
  T_{+-}^{\rm Cl} &=& - \frac{\kappa^2 P_{\Omega_0}^2}{2(2\kappa%
    P_{\Omega_0} t + A_0)}, \label{T+-:cl:com} \\
  T_{\pm\pm}^{\rm Qt} &=& -\kappa t_\pm - \kappa (P_{\Omega_0} +
    P_{\chi_0})^2, \label{T++:qt:com} \\
  T_{+-}^{\rm Qt} &=& 0.
\end{eqnarray}
Then, by using Eqs.~(\ref{rho:def}), (\ref{p:def}), (\ref{rho:non}),
and (\ref{p:non}), we obtain the energy densities and the pressures as
\begin{eqnarray}
  \rho_{\rm Cl} &=& - 3\kappa^2 P_{\Omega_0}^2 \exp \left[
    \frac{2}{\kappa}(A_0 - B_0) + 4 (P_{\Omega_0} + P_{\chi_0}^2) t
    \right], \label{rho:cl:com} \\ 
  \rho_{\rm Qt} &=& (2\kappa P_{\Omega_0} t + A_0) \exp \left[
    \frac{2}{\kappa}(A_0 - B_0) + 4 (P_{\Omega_0} + P_{\chi_0}^2) t
    \right] \nonumber \\
    & & \times \left( -2\kappa (P_{\Omega_0} + P_{\chi_0})^2 - \kappa(t_+
    + t_-) \right), \label{rho:qt:com} \\
  p_{\rm Cl} &=& - \kappa^2 P_{\Omega_0}^2 \exp \left[
    \frac{2}{\kappa}(A_0 - B_0) + 4 (P_{\Omega_0} + P_{\chi_0}^2) t
    \right], \label{p:cl:com} \\
  p_{\rm Qt} &=& (2\kappa P_{\Omega_0} t + A_0) \exp \left[
    \frac{2}{\kappa}(A_0 - B_0) + 4 (P_{\Omega_0} + P_{\chi_0}^2) t
    \right] \nonumber \\
    & & \times \left( -2\kappa (P_{\Omega_0} + P_{\chi_0})^2 - \kappa(t_+
    + t_-) \right). \label{p:qt:com}
\end{eqnarray}

Now, in order to describe the geometry from the decelerating phase to
the accelerating universe, which finally ends up with a vanishing
curvature scalar corresponding to the zero acceleration in
Ref.\cite{ky:ncdc}, one should patch two different solution at
$t=t_2$. Thus, matching the solutions in Eqs.~(\ref{sol:Omega_non})
and (\ref{sol:chi_non}) with Eqs.~(\ref{sol:Omega_com}) and
(\ref{sol:chi_com}) up to their time derivatives at $t=t_2$ yields the
following conditions:
\begin{eqnarray}
  \beta &=& \frac{P_{\Omega_0}}{2\theta_2} {\rm sech}\, (2\kappa\theta_2
            t_2), \label{patch:beta} \\
  A &=& A_0 + \frac{P_{\Omega_0}}{\theta_2} \left[ 2\kappa\theta_2 t_2 -
            \tanh (2\kappa\theta_2 t_2) \right], \label{patch:A} \\
  B &=& B_0 - \frac{P_{\chi_0}}{\theta_2} \left[ 2\kappa\theta_2 t_2 -
            \coth (2\kappa\theta_2 t_2) \right], \label{patch:B} \\
  \frac{P_{\chi_0}}{P_{\Omega_0}} &=& \tanh (2\kappa\theta_2
            t_2). \label{patch:chi} 
\end{eqnarray}
The above conditions in Eqs.~(\ref{patch:beta}) and (\ref{patch:chi}) are
rewritten as
\begin{eqnarray}
  P_{\Omega_0} &=& 2\beta\theta_2 \cosh (2\kappa\theta_2 t_2),
     \label{patch:POmega} \\
  P_{\chi_0} &=& 2\beta\theta_2 \sinh (2\kappa\theta_2 t_2).
     \label{patch:Pchi}
\end{eqnarray}
Since $da(\tau)/d\tau = d{\tilde\rho}(t)/dt$, we obtain the expanding
velocity of the universe as
\begin{equation}
  \label{vel:com}
  \frac{da(\tau)}{d\tau} = -2 (P_{\Omega_0} + P_{\chi_0}) -
  \frac{\kappa P_{\Omega_0}}{A_0 + 2\kappa P_{\Omega_0} t},
\end{equation}
which is always negative because $P_{\Omega_0} > |P_{\chi_0}|$ from
Eqs.~(\ref{patch:POmega}) and (\ref{patch:Pchi}) and $\Omega_0 = A_0 +
2\kappa P_{\Omega_0} t >0$ for any time $t$.

% curvature scalar R
\begin{figure}[t]
 \includegraphics[width=0.6\textwidth]{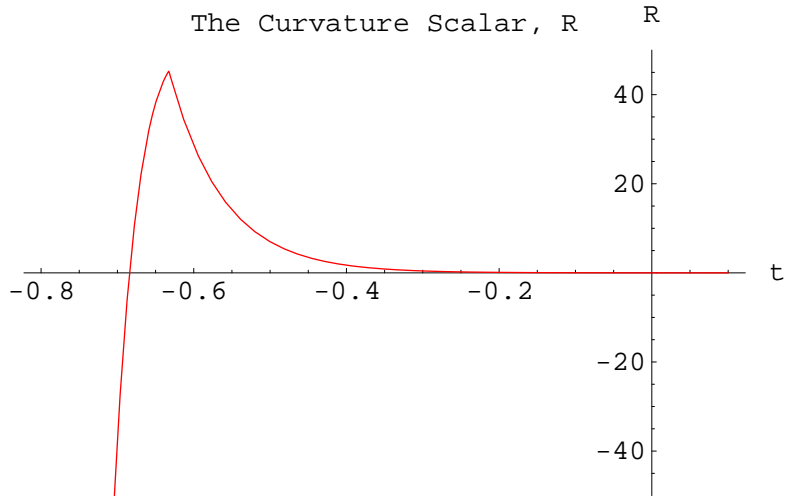}
 \caption{Change of the $\theta_2$-dependent curvature scalar in
   Eq.~(\ref{R:non}) from the negative to the positive region. After
   $t=t_2$, it is always accelerating and converges at $t \to \infty$
   by Eq.~(\ref{R:com}) for $P_{\chi_0}< 0$. This figure is plotted
   for $\beta=\kappa=\theta_2=1$, $A=10$, and $B=3$, $t>t_1$. Then,
   $t_1 \approx -1.156$ and $t_2 \approx -0.632 $. Consistent
   constants satisfying the continuity equations are chosen as
   $P_{\Omega_0} \approx 3.826$, $P_{\chi_0} \approx -3.262$, $A_0
   \approx 11.579$, and $B_0 \approx 3.301$.}
 \label{fig:R}
\end{figure}

Next, we assign one more condition of $R(t_2)=R_{\theta}(t_2)$ in
order to find the appropriate time ``$t_2$'' that connects the
respective scalar curvatures. This continuity requirement leads to
\begin{eqnarray}
 P_{\Omega_0}^2 &=& 2\beta\theta_2^2 \bigg[ \frac{2}{\kappa}
    e^{2\kappa\theta_2 t_2} \left( 2\beta\sinh (2\kappa\theta_2 t_2) 
    + A \right) \left( 2\beta\sinh (2\kappa\theta_2 t_2) + A 
    + \frac{\kappa}{2} \right) \nonumber \\
 & & - \cosh (2\kappa\theta_2 t_2) \left( 2\beta e^{2\kappa\theta_2
    t_2} + A \right)\bigg], \label{patch:R}
\end{eqnarray}
which corresponds to the requirement that derivatives up to the second
derivatives of the metric and the dilaton fields be continuous. From
the beginning, we have considered that $\kappa$, $\theta_2$, and
$P_{\Omega_0}$ to be positive and $P_{\chi_0}$ to be negative; then,
from Eq.~(\ref{patch:chi}), a consistent patching appears at the
negative value of $t_2$. Thus, we obtain the desired geometry
connecting the decelerating universe to the accelerating universe,
where its acceleration tends to vanish eventually.

% energy density
\begin{figure}[h]
 \includegraphics[width=0.6\textwidth]{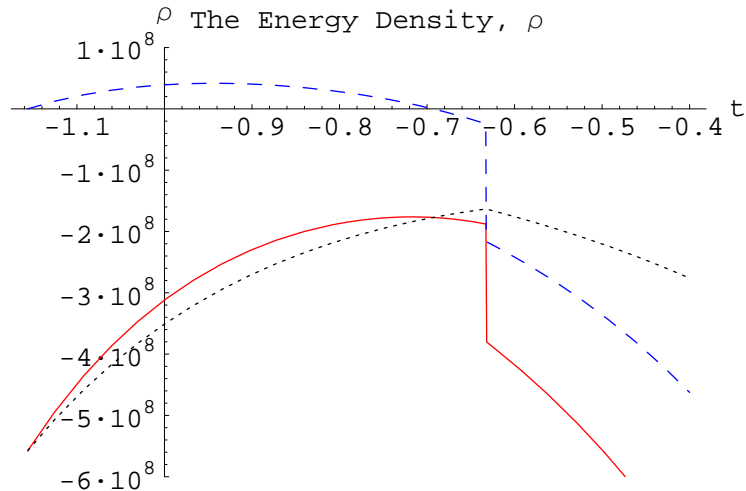}
 \caption{The solid, the dashed, and the dotted lines denote the
   energy densities of the total, the quantum, and the classical
   matter, respectively. At $t=t_2$, the total energy density is
   discontinuous. This figure is plotted for the same constants used
   in Fig.~\ref{fig:R}.} 
 \label{fig:energy} 
\end{figure}

% pressure
\begin{figure}[h]
 \includegraphics[width=0.6\textwidth]{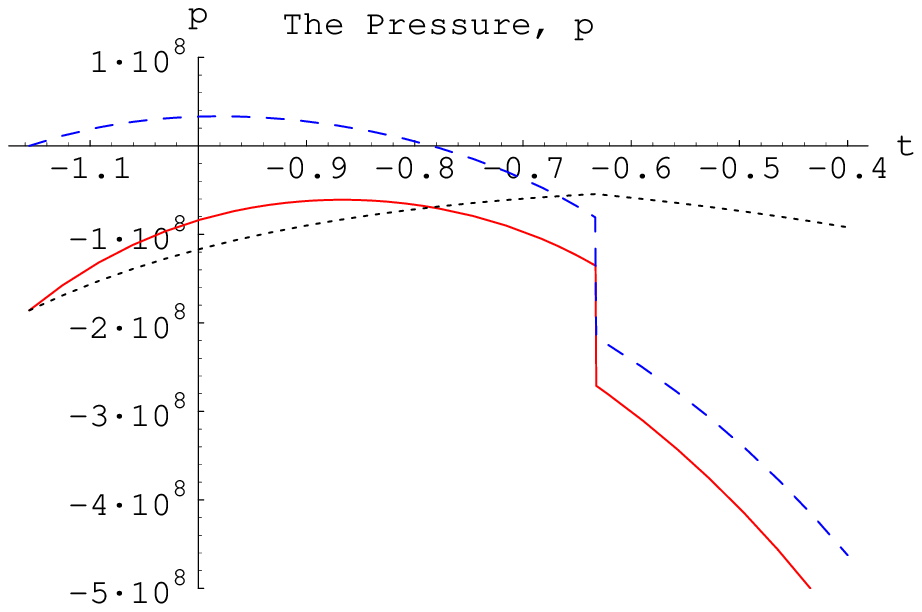}
 \caption{The solid, the dashed, and the dotted lines denote the
   pressures of the total, the quantum, and the classical matter,
   respectively. This figure is plotted for the same constants used in
   Fig.~\ref{fig:R}.}
 \label{fig:p} 
\end{figure}

\begin{figure}[h]
 \includegraphics[width=0.6\textwidth]{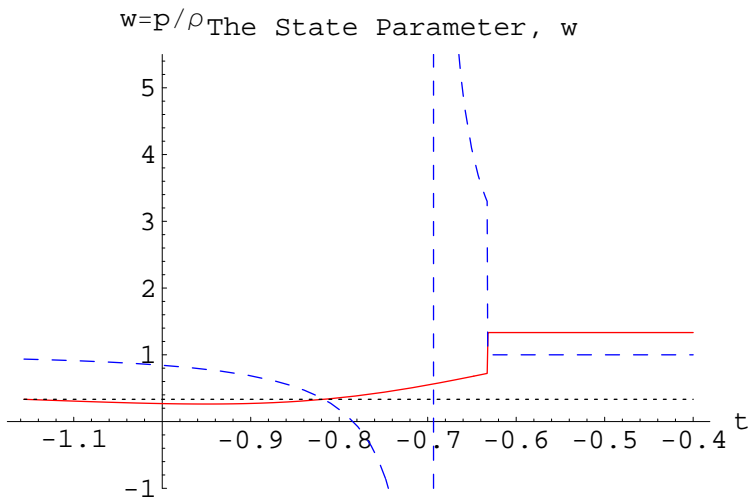}
 \caption{The solid, the dashed, and the dotted lines denote the state
   parameters of the total, the quantum, and the classical matter,
   respectively. This figure is plotted for the same constants used in
   Fig.~\ref{fig:R}.}
 \label{fig:w}
\end{figure}

Then, the behaviors of the energy densities and the pressures are
shown in Figs.~\ref{fig:energy} and \ref{fig:p}. The total energy and
pressure are negative, but the induced matter from the quantum back
reaction changes from positive energy and pressure to negative energy
and pressure. Note that $w_{\rm Cl} = p_{\rm Cl} / \rho_{\rm Cl}=1/3$
for $t>t_1$ and $w_{\rm   Qt} = p_{\rm Qt} / \rho_{\rm Qt}=1$ only for
$t>t_2$. The behavior of the state parameter $w$ is shown in
Fig.~\ref{fig:w}.

%%%%%%%%%%%%%%%%%%%%%%%%%%%%%%%%%%%%%%%%%%%%%%%%%%%%%%%%%%%%%%%%%%%%%%%

% state parameter 

%%%%%%%%%%%%%%%%%%%%%%%%%%%%%%%%%%%%%%%%%%%%%%%%%%%%%%%%%%%
%%%%%%%%%%%%               Discussion            %%%%%%%%%%
%%%%%%%%%%%%%%%%%%%%%%%%%%%%%%%%%%%%%%%%%%%%%%%%%%%%%%%%%%%
\section{Discussion}
\label{sec:dis}

In summary, the classical energy density and the classical pressure
start with a negative value that goes to the infinity so that the
equation-of-state parameter is always positive constant. As for the
quantum-mechanical energy density and pressure, they have some finite
positive and negative values alternatively in the noncommutative
region, and they jump down at the critical point $t_2$, and eventually
end up as divergent quantities in the commutative
region. Correspondingly, the state parameter is singular at the
critical time $t_2$. Except near the point, it is mostly
positive. Finally, the total energy density and the total pressure
rely on the dilatonic contribution because the classical dilaton
contribution is larger than the quantum-mechanical one, so that they
have negative values both in the noncommutative and the commutative
regions. However, the state parameter is remarkably positive;
furthermore, it is regular every time. Note that the total state
parameter from the total contributions is not the same for each
contributions from the classical and the quantum-mechanical regions
because $w  \ne w_{\rm Cl} + w_{\rm Qt}$; by definition, the exact
form is easily written as $w=(w_{\rm Cl} \rho_{\rm Cl} + w_{\rm Qt}
\rho_{\rm Qt}) / (\rho_{\rm Cl} + \rho_{\rm Qt})$, which is
reminiscent of the center-of-mass coordinate in a mechanical
system. Consequently, a positive definite value is given even though
the universe is accelerating because the quantum-mechanically-induced
energy is not positive definite.

%%%%%%%%%%%%%%%%%%%%%%%%%%%%%%%%%%%%%%%%%%%%%%%%%%%%%%%%%%%
%%%%%%%%%%%%           Acknowledgments           %%%%%%%%%%
%%%%%%%%%%%%%%%%%%%%%%%%%%%%%%%%%%%%%%%%%%%%%%%%%%%%%%%%%%%

\begin{acknowledgments}
This work was supported by the Sogang Research Grant, 20061055.  
\end{acknowledgments}

%%%%%%%%%%%%%%%%%%%%%%%%%%%%%%%%%%%%%%%%%%%%%%%%%%%%%%%%%%%%%
%%%%%%%%%%%%%%%             References       %%%%%%%%%%%%%%%%
%%%%%%%%%%%%%%%%%%%%%%%%%%%%%%%%%%%%%%%%%%%%%%%%%%%%%%%%%%%%%

\end{document}